\documentclass[%
 aip,
cp,  
 amsmath,amssymb,
 reprint,%
nofootinbib]{revtex4-2}

\usepackage{graphicx}
\usepackage{dcolumn}
\usepackage{bm}

\usepackage[utf8]{inputenc}
\usepackage[T1]{fontenc}
\usepackage{mathptmx} 
\usepackage{caption}
\usepackage{titlesec}
\usepackage[symbol]{footmisc}
\begin{document}
\title{Chameleon Mechanism in Inhomogeneous Astrophysical Objects}
\author{Noshad Khosravi Largani} 
 \email[]{corresponding author: n.khosravi@student.alzahra.ac.ir}
\author{Mohammad Taghi Mirtorabi}%
 \email[]{torabi@alzahra.ac.ir}
\affiliation{
  Department of Physics, Alzahra University,Vanak Village Street, 1993893973, Tehran, Iran 
}

\date{\today} 

\begin{abstract}
Observational evidence implying the accelerated expansion of the universe has been the motivation to develop various classes of modified gravity theories. One of them uses the so-called  "screening mechanism", which is successful in reproducing the observed gravitational behavior in large scales as well as being in agreement with tests of general relativity in the solar system. In this work, we investigate an example of scalar-tensor theories with screening mechanism, namely the profile of a Chameleon field around inhomogeneous astrophysical objects. According to \cite{Khoury:2003rn}, one can define two kinds of approaches applicable to the thin shell and thick shell regimes, that allow for a solution to the Chameleon equation of motion. For sufficiently large objects,
the scalar field can be assumed to propagate from a thin shell of the object instead of the whole body, which simplifies the problem. On the other hand, this solution is not practical in small objects. We find that in inhomogeneous objects this is not trivial and at least one more factor, which turns out to be the density, can change the way of approaching this problem. 
\end{abstract}
\maketitle
\section{Introduction} \label{introduction}
Long time has passed since Einstein's general relativity theory (GR) was introduced and since then the theory has faced several challenges. Therefore, many alternative models have been developed, mainly aimed at solving some of the problems that appear when the theory fails. For instance, \cite{Jain:2013wgs} and \cite{Nojiri:2006ri} are good references that introduce some of the modified gravity theories. One of the alternatives for a solution is adding at least one scalar field to the GR action that would form a scalar tensor theory. This scalar field's coupling to matter must be tuned to small values in order not to violate the equivalence principle and moreover it might explain some anomalous observations, like the universe accelerated expansion. The first attempt was made by Brans and Dicke~\cite{Fujii:2003pa} who tried to vary the gravitational constant that actually plays the role of the scalar field in their theory. Later on, it was suggested to add a potential in a way that the theory could be able to produce a deviation from the GR predicted behavior at large scales and at the same time to suppress the fifth force effects in the solar system, where measurements perfectly agree with GR predictions. Probing unscreened regions in order to find effects of the fifth force has been done by\cite{Sakstein:2015aac}, \cite{Sakstein:2014nfa} and many more. Here we present one of those approaches for screening mechanisms, the Chameleon field whose behavior changes depending on the environment's density. Peculiarly, we study how the Chameleon field behaves in hypothetically homogeneous stars, and in more realistic cases such as Red Giant Branch stars and White Dwarfs as two extreme cases of inhomogeneity, we find that having a thin shell solution as explained in \cite{Khoury:2003rn} that depends on the size of the objects is not always the case. The equation of motion of the Chameleon scalar field also has been studied inside and outside a spherically symmetric object in \cite{Chang:2010xh}.\\
\section{The Chameleon scalar field} \label{chameleon}
As an example of the Scalar-tensor theories, one can define a simple form of the action for the Chameleon field in Einstein frame written as:
\begin{equation}
\label{action}
 \ S=\int{d^4x \sqrt{-g}[\frac{M_{Pl}^2}{2}R-\frac{1}{2}(\partial \phi )^2-V(\phi)+L_{m}(g_{\mu \nu}^{(i)},\Psi_m^{(i)})]}
 \end{equation}
 where $\phi$ represents the Chameleon scalar field, $\Psi_m^{(i)}$ are matter fields, and 
 \begin{equation}
 \ g_{\mu \nu}^{(i)} = e^{2\beta _i \phi /M_{pl}} g_{\mu \nu}
 \end{equation}
 is the metric that describes the geodesics for the i-th matter field. $\beta_i$ are dimensionless coupling constants that can vary for different kinds of matter, but in our case we assume them to be of order unity since we know that gravity behaves the same for all of them. Therefore, the (i) indices can be omitted.
 Varying the action (\ref{action}) with respect to $\phi$ gives the Chameleon equation of motion:
 \begin{equation}
 \label{cham}
 \Box \phi= V_{,\phi}+\frac{\beta}{M_{pl}}e^{4\beta \phi /M_{pl}} g_{(i)}^{\mu \nu} T_{\mu \nu}^{(i)}
 \end{equation}
Moreover, we consider solving this equation for non-relativistic matter.  Consequently,  the trace of energy-momentum tensor represents the energy density in Jordan frame:
  \begin{equation}
 \  g_{(i)}^{\mu \nu} T_{\mu \nu}^{(i)} =  \tilde{\rho}_{(i)} 
 \end{equation} 
 We employ a conformal translation as it is simpler to solve the equation in Einstein's frame and obtain:
  \begin{equation}
  \rho^{(i)}\equiv e^{3\beta \phi / M_{pl}} \tilde{\rho}_{(i)}
  \end{equation}
 so that equation~(\ref{cham}) transforms into 
 \begin{equation}
 \label{Eom}
 \Box \phi= V_{,\phi}+\frac{\beta}{M_{pl}}e^{\beta  \phi /M_{pl}} \rho.
\end{equation}

In this equation of motion there shall appear two important mass terms.
The first one is a runaway potential which is a monotonic decreasing function, and the second is deduced from the conformal coupling of matter fields, $e^{\beta \phi / M_{pl}}$. Combining these two terms gives us an effective potential whose derivative with respect to $\phi$ has appeared in the above equation and plays an important role in the Chameleon model:
\begin{equation}
\label{Veff}
 \ V_{eff}(\phi) = V(\phi)+ \rho e^{\beta \phi /M_{pl}}.
\end{equation}
For the first term on the left we consider inverse power-law potentials of the form
\begin{equation}
\ V(\phi)= \frac{M^{4+n}}{\phi^n}
\end{equation}
in which $n$ is a positive constant and $M$ has units of mass. Since the field values divided by the Planck mass do not contribute, we can 
use a linear approximation
\begin{equation}
\ e^{\beta_i \phi/M_{pl}} \simeq 1+\beta \phi/M_{pl}
\end{equation}
so that the equation (\ref{Veff}) can be written as:
\begin{equation}
\ V_{eff}(\phi)= \frac{M^{4+n}}{\phi^n}+ \rho +\beta \rho \phi/M_{pl}.
\end{equation}

Combination of inverse power-law potential and the exponential term
provides the effective potential with a minimum. Hence, depending on how these two terms change, this minimum can change.
In other words, both the minimum of the field $\phi_{min}$ and it's mass $m_{min}$ depend on local matter density. 
This means that in places with high density, $\phi_{min}$ shall be smaller and by decreasing density, $\phi_{min}$ will change to greater values~\cite{Joyce:2014kja}.  
Differentiating the effective potential with respect to the scalar field gives the minimum value:
\begin{equation}
\label{PhiMin}
\phi_{min}= [{\frac{nM_{pl}M^{4+n}}{\beta\rho}}]^{\frac{1}{n+1}}.
\end{equation}
Moreover, the second derivative of $V_{eff}$ with respect to $\phi$ can define the mass around the minimum of the effective potential, which takes the form:
\begin{equation}
\label{meff}
\frac{d^2V_{eff}}{d\phi^2}=n(n+1) \frac{M^{4+n}}{\phi_{min}^{n+2}}\frac{\beta^2 \rho}{M_{pl}^2}\equiv m_{eff}^2.
\end{equation}
Substituting equation (\ref{PhiMin}) into (\ref{meff}) gives:
\begin{equation}
\label{meffDeriv}
m_{eff}^2\equiv n(n+1) M^{-(\frac{4+n}{n+1})}(\frac{\beta\rho}{nM_{pl}})^{\frac{n+2}{n+1}}
\end{equation}
As it is visible in equation (\ref{meffDeriv}), the mass of the scalar field is directly related to density.
 This means that increasing local density will increase the field mass and hence its range will scale down. In particular, the Chameleon mechanism is a specific way of screening mechanism and is able to suppress the fifth force in the solar system. But as it is expected, there might be places with low densities that would reveal the effect of fifth force in them.
The Chameleon mechanism usually works in objects with a "thin shell". A body has a thin shell if the Chameleon field inside it is constant everywhere but in a small region close to the surface of the object.
 This assumption would make the effect of the field weaker in large objects even if the field's range is considerable.

 If the object is sufficiently small, the thin shell regime will not be valid and the Chameleon approach is within the so called "thick shell" regime~\cite{Khoury:2003rn}. This issue is found to be trivial for homogeneous profiles, but we claim that it might not be trivial for some astrophysical objects that have significant inhomogeneous density profiles.

\section{The numerical solution of Chameleon equation of motion} \label{solution}
The main aim of this work is to investigate how the Chameleon field behaves in astrophysical objects, which are assumed to have inhomogeneous density distributions. In order to investigate the Chameleon effects, we first consider a homogeneous profile that will allow for comparison with inhomogeneous ones. The first density profile is simply distance dependent, followed by two other more realistic profiles that describe a white dwarf and an RGB star, respectively. 
In order to perform computations, we have written all the equations in dimensionless form and used the relaxation method described in~\cite{recepie}.
\vspace{-0.25cm}
\subsection{Homogeneous profile}
The Homogeneous energy density profile we assume is
\begin{equation}
\rho_0(r)=
 \left\{ \begin{array}{ll}
\rho_c \qquad r<{R_c} ,  \\
\rho_G \qquad r>{R_c} .\end{array} \right.    
\end{equation}
where $r$ is the distance from the center, $R_c$ is the total radius, $\rho_c$ is the constant density value in center of the object, and $\rho_G$ is the background density.
In order to make the dimensionless form of equation (\ref{Eom}) using a spherical symmetry, one can rescale the scalar field with respect to the value of the field at the minimum effective potential by defining $\varphi \equiv \phi/{\phi_c}$ and the same procedure for the distance from the center of the object $x \equiv r/{R_c}$ and also defining $m_c$ which is the effective mass at the minimum effective potential. After some algebraic manipulations, The equation (\ref{Eom}) can be written as
\begin{equation}
\frac{d^2\varphi}{dx^2}+ \frac{2}{x} \frac{d\varphi}{dx}= \frac{(m_cR_c)^2}{n+1}[\frac{\rho_0(x)}{\rho_c}-\frac{1}{\varphi^{n+1}}].
\end{equation}
We treat the above equation as a boundary value problem with the following boundary conditions:
\begin{equation}
 \left\{ \begin{array}{ll}
\frac{d\phi}{dx}(x=0)=0  \\
\phi(x\rightarrow\infty) = \phi_G\end{array} \right.  
\end{equation}
in which $({\phi_G}/{\phi_c}) $ is the dimensionless scalar field in the galaxy, and it is proportional to ${\rho_G}/{\rho_c}= 5\times10^{4}$\cite{Tsujikawa:2009yf}. This equation has also been solved numerically in~\cite{Silvestri:2011ch} .
\subsection{Inhomogeneous profiles}
\begin{figure}[t]
	\centering
		\includegraphics[width=0.6\textwidth]{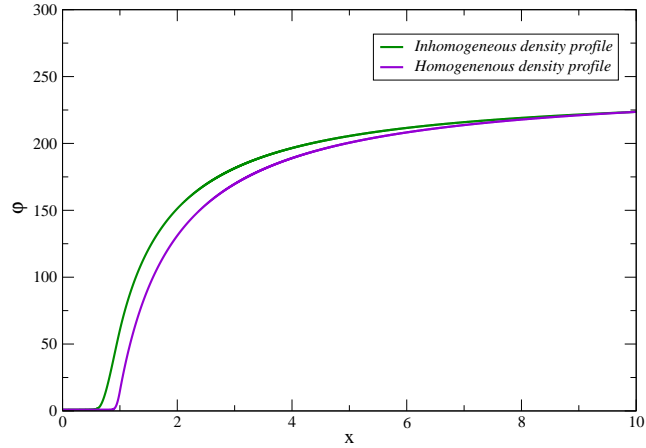}
	\caption{Chameleon field behavior as a function of distance from the center of the object in an inhomogeneous profile (green) in comparison to a homogeneous density profile (purple).}
	\label{densities}
\end{figure} 

\begin{figure}
\centering
\begin{minipage}{.5\textwidth}
  \centering
  \includegraphics[width=.98\linewidth]{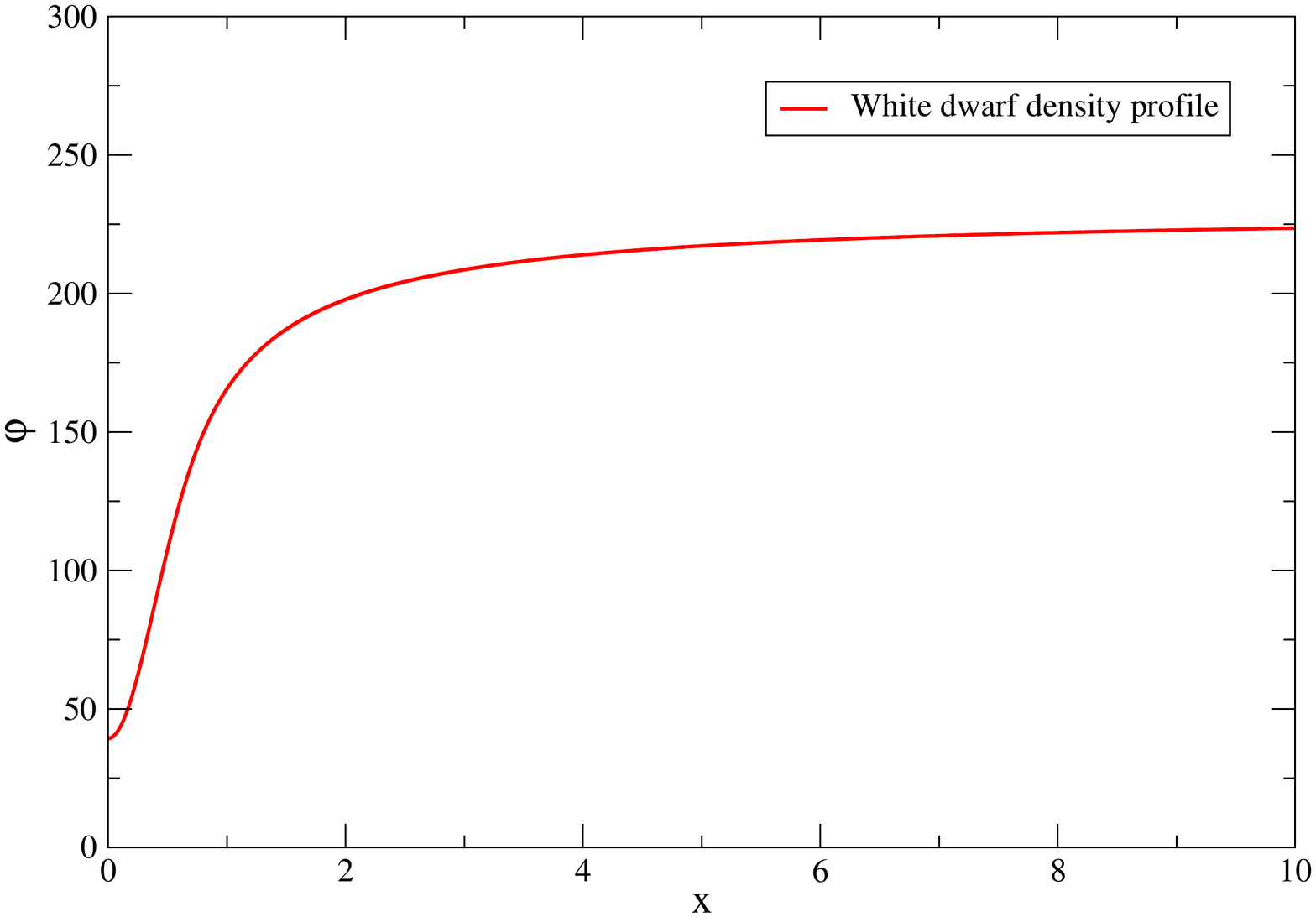}\\
  \captionof{figure}{Chameleon field behavior as a function of distance from the center of the object in a white dwarf density profile.}
  \label{wd}
\end{minipage}%
\begin{minipage}{.5\textwidth}
  \centering
  \includegraphics[width=.98\linewidth]{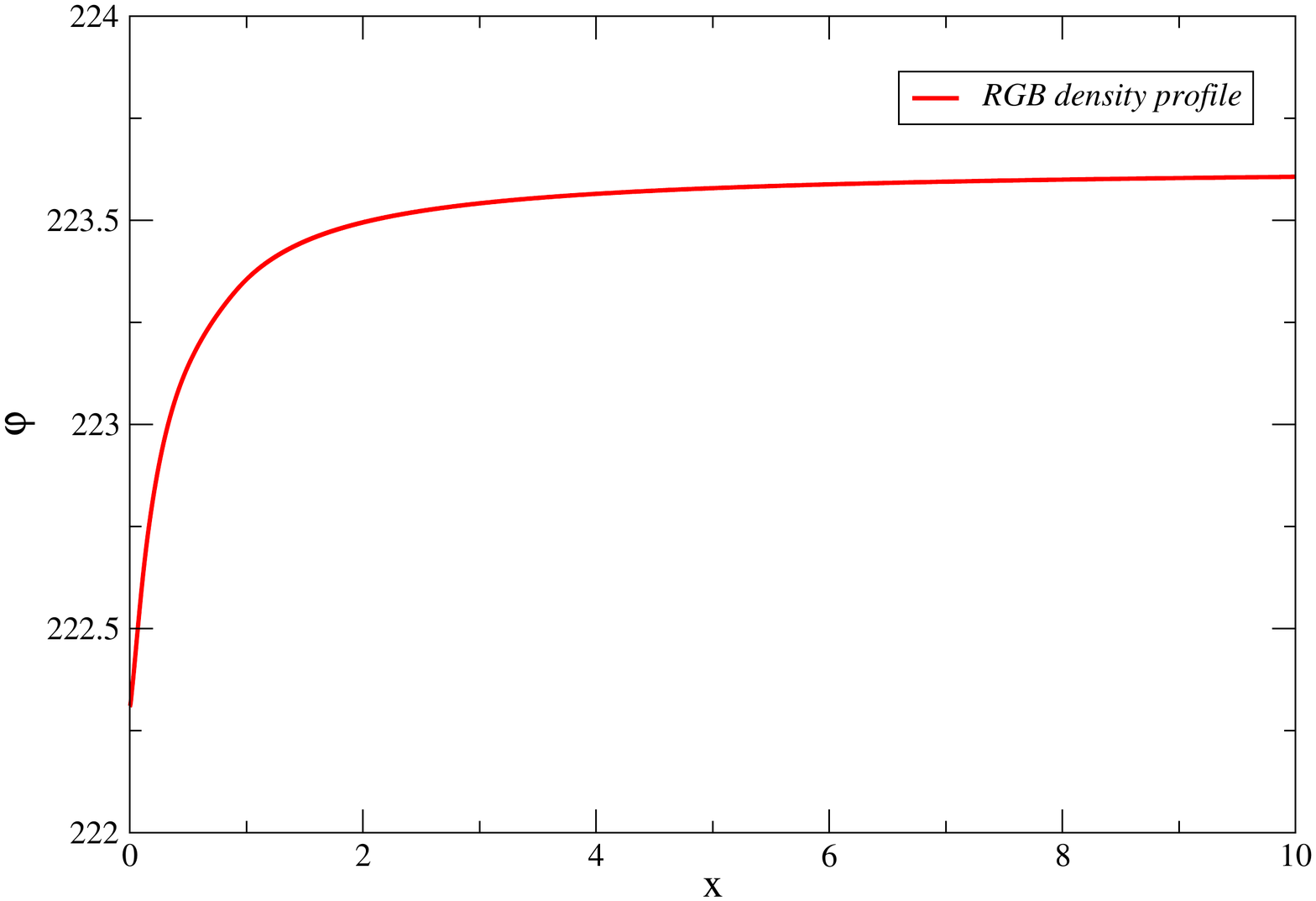}\\
  \captionof{figure}{Chameleon field behavior as a function of distance from the center of the object in a RGB star density profile.}
  \label{rgb}
\end{minipage}
\end{figure}

In order to solve a more realistic problem, we can consider a density profile inside the object as a function of the dimensionless distance:
\begin{equation}
\rho(x)=
\left \{  \begin{array}{ll}
1-x^2 \qquad x<1   \\
\rho_G/ \rho_c \qquad x>1\end{array} \right.   
\end{equation}
The two density profiles related to a white dwarf and a red giant were produced using the MESA simulation code. Modules for Experiments in Stellar Astrophysics MESA is a general, modern stellar evolution code able to run in a wide range of environments \cite{Paxton:2013pj}. 
\section{Results} \label{results}
As it was expected in a thin shell regime, the scalar field is suppressed inside the object ($x<1$) and only in a very small region (the thin shell) it could start to increase to finally reach to it's maximum outside the object, the purple line in Fig.~\ref{densities}.
By varying the density, as it is shown in the green curve of Fig.~\ref{densities}, the region of thin shell increases, however we still can have the thin shell regime for this configuration.
Fig.~\ref{wd} shows the profile of a white dwarf obtained from the MESA code. As it is shown there, the Chameleon field is not suppressed even inside the object. This simply means that the thin shell regime is not valid anymore therefore the Chameleon field permeates the entire object.
As it is well known, in comparison to other astrophysical objects, White Dwarfs count as small, compact stars and consequently the thin shell regime applies.
On the other hand, this means that if we choose a big configuration, the thin shell will still be visible. Fig.~\ref{rgb} shows the profile of an RGB star. The first conclusion is that the behavior of the field is almost the same as a white dwarf, without a thin shell. Therefore, size is not the only factor for having a thin shell regime, but density.
Furthermore, a comparison of figures \ref{wd} and \ref{rgb} shows that the minimum value of the scalar field in the RGB star is considerably higher than the one of a white dwarf. This is in agreement with the behavior of Chameleon: the lower the density, the higher the minimum value of the scalar field.

\section{Outlook and conclusions}
In Sec. \ref{introduction} and \ref{chameleon} we have explained a special way of modifying general gravity, employing a scalar-tensor theory. We also discussed a type of modified gravity approach that uses a mechanism that although it satisfies solar system tests of GR by suppressing the fifth force, it may bear significant effects in larger scales of the universe.
One of the simple ways to introduce a screening mechanism is using the prescription of the Chameleon model, used here. In Sec. \ref{solution} we have explained more details of the resulting numerical solutions and finally in Sec. \ref{results}, we showed the concluding results of the simple practice carried out in this work.
We have found that in addition to size, varying density can also result into effects that fall under the thin shell or thick shell regime. As a following step, inclusion of a time dependent source shall be considered to investigate the behavior of the time-dependent Chameleon field as well as the effects of inhomogeneous environments.

\begin{acknowledgments}
The authors would like to express their appreciation to the organizers of the AYSS 2019 conference for their hospitality. We are grateful to Y. Akrami for scientific collaboration and helpful suggestions. N. K-L. acknowledges support from D. Blaschke  for participating in AYSS conference 2019 and is thankful to H. Bagheri for assistance with computations.

\end{acknowledgments}
\vspace{-1cm}
\bibliography{chameleon}

\begin{thebibliography}{12}%
\makeatletter
\providecommand \@ifxundefined [1]{%
 \@ifx{#1\undefined}
}%
\providecommand \@ifnum [1]{%
 \ifnum #1\expandafter \@firstoftwo
 \else \expandafter \@secondoftwo
 \fi
}%
\providecommand \@ifx [1]{%
 \ifx #1\expandafter \@firstoftwo
 \else \expandafter \@secondoftwo
 \fi
}%
\providecommand \natexlab [1]{#1}%
\providecommand \enquote  [1]{``#1''}%
\providecommand \bibnamefont  [1]{#1}%
\providecommand \bibfnamefont [1]{#1}%
\providecommand \citenamefont [1]{#1}%
\providecommand \href@noop [0]{\@secondoftwo}%
\providecommand \href [0]{\begingroup \@sanitize@url \@href}%
\providecommand \@href[1]{\@@startlink{#1}\@@href}%
\providecommand \@@href[1]{\endgroup#1\@@endlink}%
\providecommand \@sanitize@url [0]{\catcode `\\12\catcode `\$12\catcode
  `\&12\catcode `\#12\catcode `\^12\catcode `\_12\catcode `\%12\relax}%
\providecommand \@@startlink[1]{}%
\providecommand \@@endlink[0]{}%
\providecommand \url  [0]{\begingroup\@sanitize@url \@url }%
\providecommand \@url [1]{\endgroup\@href {#1}{\urlprefix }}%
\providecommand \urlprefix  [0]{URL }%
\providecommand \Eprint [0]{\href }%
\providecommand \doibase [0]{http://dx.doi.org/}%
\providecommand \selectlanguage [0]{\@gobble}%
\providecommand \bibinfo  [0]{\@secondoftwo}%
\providecommand \bibfield  [0]{\@secondoftwo}%
\providecommand \translation [1]{[#1]}%
\providecommand \BibitemOpen [0]{}%
\providecommand \bibitemStop [0]{}%
\providecommand \bibitemNoStop [0]{.\EOS\space}%
\providecommand \EOS [0]{\spacefactor3000\relax}%
\providecommand \BibitemShut  [1]{\csname bibitem#1\endcsname}%
\let\auto@bib@innerbib\@empty
\bibitem [{\citenamefont {Khoury}\ and\ \citenamefont
  {Weltman}(2004)}]{Khoury:2003rn}%
  \BibitemOpen
  \bibfield  {author} {\bibinfo {author} {\bibfnamefont {J.}~\bibnamefont
  {Khoury}}\ and\ \bibinfo {author} {\bibfnamefont {A.}~\bibnamefont
  {Weltman}},\ }\bibfield  {title} {\enquote {\bibinfo {title} {{Chameleon
  cosmology}},}\ }\href {\doibase 10.1103/PhysRevD.69.044026} {\bibfield
  {journal} {\bibinfo  {journal} {Phys. Rev.}\ }\textbf {\bibinfo {volume}
  {D69}},\ \bibinfo {pages} {044026} (\bibinfo {year} {2004})},\ \Eprint
  {http://arxiv.org/abs/astro-ph/0309411} {arXiv:astro-ph/0309411 [astro-ph]}
  \BibitemShut {NoStop}%
\bibitem [{\citenamefont {Jain}\ \emph {et~al.}(2013)\citenamefont {Jain} \emph
  {et~al.}}]{Jain:2013wgs}%
  \BibitemOpen
  \bibfield  {author} {\bibinfo {author} {\bibfnamefont {B.}~\bibnamefont
  {Jain}} \emph {et~al.},\ }\bibfield  {title} {\enquote {\bibinfo {title}
  {{Novel Probes of Gravity and Dark Energy}},}\ }\href@noop {} {\  (\bibinfo
  {year} {2013})},\ \Eprint {http://arxiv.org/abs/1309.5389} {arXiv:1309.5389
  [astro-ph.CO]} \BibitemShut {NoStop}%
\bibitem [{\citenamefont {Nojiri}\ and\ \citenamefont
  {Odintsov}(2006)}]{Nojiri:2006ri}%
  \BibitemOpen
  \bibfield  {author} {\bibinfo {author} {\bibfnamefont {S.}~\bibnamefont
  {Nojiri}}\ and\ \bibinfo {author} {\bibfnamefont {S.~D.}\ \bibnamefont
  {Odintsov}},\ }\bibfield  {title} {\enquote {\bibinfo {title} {{Introduction
  to modified gravity and gravitational alternative for dark energy.
  Theoretical physics: Current mathematical topics in gravitation and
  cosmology. Proceedings, 42nd Karpacz Winter School, Ladek, Poland, February
  6-11, 2006}},}\ }\href {\doibase 10.1142/S0219887807001928} {\bibfield
  {journal} {\bibinfo  {journal} {eConf}\ }\textbf {\bibinfo {volume}
  {C0602061}},\ \bibinfo {pages} {06} (\bibinfo {year} {2006})},\ \bibinfo
  {note} {[Int. J. Geom. Meth. Mod. Phys.4,115(2007)]},\ \Eprint
  {http://arxiv.org/abs/hep-th/0601213} {arXiv:hep-th/0601213 [hep-th]}
  \BibitemShut {NoStop}%
\bibitem [{\citenamefont {Fujii}\ and\ \citenamefont
  {Maeda}(2007)}]{Fujii:2003pa}%
  \BibitemOpen
  \bibfield  {author} {\bibinfo {author} {\bibfnamefont {Y.}~\bibnamefont
  {Fujii}}\ and\ \bibinfo {author} {\bibfnamefont {K.}~\bibnamefont {Maeda}},\
  }\bibfield  {title} {\enquote {\bibinfo {title} {The scalar-tensor theory of
  gravitation},}\ }\href
  {http://www.cambridge.org/uk/catalogue/catalogue.asp?isbn=0521811597} {\
  \bibinfo {series} {Cambridge Monographs on Mathematical Physics} (\bibinfo
  {year} {2007})}\BibitemShut {NoStop}%
\bibitem [{\citenamefont {Sakstein}(2015)}]{Sakstein:2015aac}%
  \BibitemOpen
  \bibfield  {author} {\bibinfo {author} {\bibfnamefont {J.}~\bibnamefont
  {Sakstein}},\ }\bibfield  {title} {\enquote {\bibinfo {title} {{Testing
  Gravity Using Dwarf Stars}},}\ }\href {\doibase 10.1103/PhysRevD.92.124045}
  {\bibfield  {journal} {\bibinfo  {journal} {Phys. Rev.}\ }\textbf {\bibinfo
  {volume} {D92}},\ \bibinfo {pages} {124045} (\bibinfo {year} {2015})},\
  \Eprint {http://arxiv.org/abs/1511.01685} {arXiv:1511.01685 [astro-ph.CO]}
  \BibitemShut {NoStop}%
\bibitem [{\citenamefont {Sakstein}, \citenamefont {Jain},\ and\ \citenamefont
  {Vikram}(2014)}]{Sakstein:2014nfa}%
  \BibitemOpen
  \bibfield  {author} {\bibinfo {author} {\bibfnamefont {J.}~\bibnamefont
  {Sakstein}}, \bibinfo {author} {\bibfnamefont {B.}~\bibnamefont {Jain}}, \
  and\ \bibinfo {author} {\bibfnamefont {V.}~\bibnamefont {Vikram}},\
  }\bibfield  {title} {\enquote {\bibinfo {title} {{Detecting modified gravity
  in the stars}},}\ }\href {\doibase 10.1142/S0218271814420024} {\bibfield
  {journal} {\bibinfo  {journal} {Int. J. Mod. Phys.}\ }\textbf {\bibinfo
  {volume} {D23}},\ \bibinfo {pages} {1442002} (\bibinfo {year} {2014})},\
  \Eprint {http://arxiv.org/abs/1409.3708} {arXiv:1409.3708 [astro-ph.CO]}
  \BibitemShut {NoStop}%
\bibitem [{\citenamefont {Chang}\ and\ \citenamefont
  {Hui}(2011)}]{Chang:2010xh}%
  \BibitemOpen
  \bibfield  {author} {\bibinfo {author} {\bibfnamefont {P.}~\bibnamefont
  {Chang}}\ and\ \bibinfo {author} {\bibfnamefont {L.}~\bibnamefont {Hui}},\
  }\bibfield  {title} {\enquote {\bibinfo {title} {{Stellar Structure and Tests
  of Modified Gravity}},}\ }\href {\doibase 10.1088/0004-637X/732/1/25}
  {\bibfield  {journal} {\bibinfo  {journal} {Astrophys. J.}\ }\textbf
  {\bibinfo {volume} {732}},\ \bibinfo {pages} {25} (\bibinfo {year} {2011})},\
  \Eprint {http://arxiv.org/abs/1011.4107} {arXiv:1011.4107 [astro-ph.CO]}
  \BibitemShut {NoStop}%
\bibitem [{\citenamefont {Joyce}\ \emph {et~al.}(2015)\citenamefont {Joyce},
  \citenamefont {Jain}, \citenamefont {Khoury},\ and\ \citenamefont
  {Trodden}}]{Joyce:2014kja}%
  \BibitemOpen
  \bibfield  {author} {\bibinfo {author} {\bibfnamefont {A.}~\bibnamefont
  {Joyce}}, \bibinfo {author} {\bibfnamefont {B.}~\bibnamefont {Jain}},
  \bibinfo {author} {\bibfnamefont {J.}~\bibnamefont {Khoury}}, \ and\ \bibinfo
  {author} {\bibfnamefont {M.}~\bibnamefont {Trodden}},\ }\bibfield  {title}
  {\enquote {\bibinfo {title} {{Beyond the Cosmological Standard Model}},}\
  }\href {\doibase 10.1016/j.physrep.2014.12.002} {\bibfield  {journal}
  {\bibinfo  {journal} {Phys. Rept.}\ }\textbf {\bibinfo {volume} {568}},\
  \bibinfo {pages} {1--98} (\bibinfo {year} {2015})},\ \Eprint
  {http://arxiv.org/abs/1407.0059} {arXiv:1407.0059 [astro-ph.CO]} \BibitemShut
  {NoStop}%
\bibitem [{\citenamefont {Press}\ \emph {et~al.}(1992)\citenamefont {Press},
  \citenamefont {Flannery}, \citenamefont {Teukolsky},\ and\ \citenamefont
  {Vetterling}}]{recepie}%
  \BibitemOpen
  \bibfield  {author} {\bibinfo {author} {\bibfnamefont {W.}~\bibnamefont
  {Press}}, \bibinfo {author} {\bibfnamefont {B.}~\bibnamefont {Flannery}},
  \bibinfo {author} {\bibfnamefont {S.}~\bibnamefont {Teukolsky}}, \ and\
  \bibinfo {author} {\bibfnamefont {W.}~\bibnamefont {Vetterling}},\ }\bibfield
   {title} {\enquote {\bibinfo {title} {Numerical recipes: The art of
  scientific computing},}\ }\href@noop {} {\  (\bibinfo {year}
  {1992})}\BibitemShut {NoStop}%
\bibitem [{\citenamefont {Tsujikawa}, \citenamefont {Tamaki},\ and\
  \citenamefont {Tavakol}(2009)}]{Tsujikawa:2009yf}%
  \BibitemOpen
  \bibfield  {author} {\bibinfo {author} {\bibfnamefont {S.}~\bibnamefont
  {Tsujikawa}}, \bibinfo {author} {\bibfnamefont {T.}~\bibnamefont {Tamaki}}, \
  and\ \bibinfo {author} {\bibfnamefont {R.}~\bibnamefont {Tavakol}},\
  }\bibfield  {title} {\enquote {\bibinfo {title} {{Chameleon scalar fields in
  relativistic gravitational backgrounds}},}\ }\href {\doibase
  10.1088/1475-7516/2009/05/020} {\bibfield  {journal} {\bibinfo  {journal}
  {JCAP}\ }\textbf {\bibinfo {volume} {0905}},\ \bibinfo {pages} {020}
  (\bibinfo {year} {2009})},\ \Eprint {http://arxiv.org/abs/0901.3226}
  {arXiv:0901.3226 [gr-qc]} \BibitemShut {NoStop}%
\bibitem [{\citenamefont {Silvestri}(2011)}]{Silvestri:2011ch}%
  \BibitemOpen
  \bibfield  {author} {\bibinfo {author} {\bibfnamefont {A.}~\bibnamefont
  {Silvestri}},\ }\bibfield  {title} {\enquote {\bibinfo {title} {{Scalar
  radiation from Chameleon-shielded regions}},}\ }\href {\doibase
  10.1103/PhysRevLett.106.251101} {\bibfield  {journal} {\bibinfo  {journal}
  {Phys. Rev. Lett.}\ }\textbf {\bibinfo {volume} {106}},\ \bibinfo {pages}
  {251101} (\bibinfo {year} {2011})},\ \Eprint {http://arxiv.org/abs/1103.4013}
  {arXiv:1103.4013 [astro-ph.CO]} \BibitemShut {NoStop}%
\bibitem [{\citenamefont {{Paxton}}\ \emph {et~al.}(2013)\citenamefont
  {{Paxton}}, \citenamefont {{Cantiello}}, \citenamefont {{Arras}},
  \citenamefont {{Bildsten}}, \citenamefont {{Brown}}, \citenamefont
  {{Dotter}}, \citenamefont {{Mankovich}}, \citenamefont {{Montgomery}},
  \citenamefont {{Stello}},\ and\ \citenamefont {{Timmes}}}]{Paxton:2013pj}%
  \BibitemOpen
  \bibfield  {author} {\bibinfo {author} {\bibfnamefont {B.}~\bibnamefont
  {{Paxton}}}, \bibinfo {author} {\bibfnamefont {M.}~\bibnamefont
  {{Cantiello}}}, \bibinfo {author} {\bibfnamefont {P.}~\bibnamefont
  {{Arras}}}, \bibinfo {author} {\bibfnamefont {L.}~\bibnamefont {{Bildsten}}},
  \bibinfo {author} {\bibfnamefont {E.~F.}\ \bibnamefont {{Brown}}}, \bibinfo
  {author} {\bibfnamefont {A.}~\bibnamefont {{Dotter}}}, \bibinfo {author}
  {\bibfnamefont {C.}~\bibnamefont {{Mankovich}}}, \bibinfo {author}
  {\bibfnamefont {M.~H.}\ \bibnamefont {{Montgomery}}}, \bibinfo {author}
  {\bibfnamefont {D.}~\bibnamefont {{Stello}}}, \ and\ \bibinfo {author}
  {\bibfnamefont {F.~X.}\ \bibnamefont {{Timmes}}},\ }\bibfield  {title}
  {\enquote {\bibinfo {title} {{Modules for Experiments in Stellar Astrophysics
  (MESA): Planets, Oscillations, Rotation, and Massive Stars}},}\ }\href
  {\doibase 10.1088/0067-0049/208/1/4} {\ \textbf {\bibinfo {volume} {208}},\
  \bibinfo {eid} {4} (\bibinfo {year} {2013})},\ \Eprint
  {http://arxiv.org/abs/1301.0319} {arXiv:1301.0319 [astro-ph.SR]} \BibitemShut
  {NoStop}%
\end{thebibliography}%
%
\end{document}